\documentclass[aps, prl, preprint]{revtex4-1}

\usepackage{graphicx}
\usepackage{amsmath}
\usepackage{amssymb}
\usepackage{color}

\begin{document}

\title{Adjoint Fokker-Planck equation and runaway electron dynamics}

\author{Chang Liu}
\affiliation{Princeton University, Princeton, New Jersey 08544, USA}
\author{Dylan P. Brennan}
\affiliation{Princeton University, Princeton, New Jersey 08544, USA}
\author{Allen H. Boozer}
\affiliation{Columbia University, New York, New York 10027, USA}
\author{Amitava Bhattacharjee}
\affiliation{Princeton University, Princeton, New Jersey 08544, USA}

\date{\today}

\begin{abstract}
The adjoint Fokker-Planck equation method is applied to study the runaway probability function and the expected slowing-down time for highly relativistic runaway electrons, including the loss of energy due to synchrotron radiation.
{In direct correspondence to Monte Carlo simulation methods,} the runaway probability function has a smooth transition {across} the runaway separatrix, which can be attributed to effect of the pitch angle scattering term in the kinetic equation. {However, for the same numerical accuracy, the adjoint method is more efficient than the Monte Carlo method.}  The expected slowing-down time gives a {novel} method to estimate the runaway current decay time in experiments. {A new result from this} {work} {is that the decay rate of high energy electrons is very slow when $E$ is close to the critical electric field.} This {effect contributes further to a hysteresis previously found} in the runaway electron population.
\end{abstract}

\maketitle
 
It is well known that under an external electric field a superthermal electron in plasma can run away from the bounds of the collisional force and get accelerated to very high energy\cite{dreicer_electron_1959}.
Runaway electrons can be produced in tokamak disruptions, which is an important issue for disruption mitigation on the International Thermonuclear Experimental Reactor (ITER). 
Further studies have thus been motivated to address the dynamics of the runaway electrons in momentum space\cite{connor_relativistic_1975,rosenbluth_theory_1997,boozer_theory_2015}.
Recently it has been shown that in the presence of a magnetic field, the synchrotron radiation reaction force can play an important role in the runaway electron dynamics\cite{martin-solis_effect_2000,martin-solis_experimental_2010,stahl_synchrotron_2013,aleynikov_theory_2015,boozer_theory_2015,hirvijoki_radiation_2015}. Combined with pitch angle scattering, the radiation force can produce additional stopping power\cite{martin-solis_experimental_2010}. The resulting effects include increase of the critical electric field $E_{0}$ (which will be above the Connor-Hastie field $E_{c}$)\cite{martin-solis_effect_2000,aleynikov_theory_2015}, and modification of the runaway electron growth and decay rate. In addition, the stopping power can help form an ``attractor'' in electron momentum space, which can lead to a bump-on-tail distribution\cite{hirvijoki_radiation_2015}. Taking these into account, simulations\cite{stahl_effective_2015} have produced results that qualitatively agree with experiment\cite{paz-soldan_growth_2014} { for the electric field} { above which x-ray signals indicate a runaway population is sustained}.
It is believed that other effects like bremsstrahlung radiation and magnetic fluctuations can also influence the runaway growth and decay. To better understand these effects it is very important to develop theoretical tools that complement numerical simulations and can provide deeper physical insight into the phase-space structure of runaway electrons.

In this paper we study the runaway electron dynamics in momentum space by solving the adjoint Fokker-Planck equation (FPE) in momentum space, which is a general method that offers significant conceptual and computational advantages. The homogeneous adjoint FPE was first introduced to study the first passage problem\cite{siegert_first_1951}. It has been applied to calculate the neutron generation probability\cite{bell_stochastic_1965}, the response function of the current drive\cite{fisch_transport_1986}, and the runaway probability\cite{karney_current_1986}.
Here we demonstrate that the adjoint FPE can be used not only to study the probability function and its moments, as is often done, but also to calculate subtler and experimentally relevant quantities like the slowing-down time for existing superthermal electrons using the nonhomogeneous form of the FPE (see Appendix \ref{adjoint-equatin}). This method takes into account all the terms in the kinetic equation, and improves upon the test particle method which ignores the diffusion term\cite{parks_avalanche_1999,martin-solis_momentumspace_1998}.
{ In addition, the adjoint method is much more efficient than the Monte Carlo method since it can provide detailed information in all of momentum space by solving a single partial differential equation (PDE) once. It can be more advantageous to study certain physical effects using the adjoint method, because the adjoint FPE has a direct relation to the standard Fokker-Planck equation.} 

We demonstrate that the the runaway probability function shows a transition layer in momentum space, which agrees with the separatrix found by the test particle method.
However, due to the effect of pitch angle scattering, the layer of finite width provides a smooth transition rather than a discontinuous transition represented by a step function (in momentum space).
The expected slowing-down time we calculate characterizes the runaway electron beam decay time, which gives a new perspective to the study of runaway current decay in both the quiescent runaway electron (QRE)\cite{paz-soldan_growth_2014} and the plateau\cite{andersson_damping_2001} regimes. The result shows that the electric field { must be well below $E_{0}$} for significant decay {to occur}.

{In the established model of} runaway electron dynamics{, w}hen $E$ is larger than the critical electric field and the radiation effect is weak, electrons initially in the high energy regime can continue to be accelerated and run away. On the other hand, electrons initially in the low energy regime will be decelerated and fall back into the Maxwellian population. Thus the destinations of electrons in the long time limit depend on their initial momentum. The radiation force can be an additional source of stopping power, but it can only dominate the electric force in the very high energy regime when $E$  is much larger than the critical electric field. The kinetic equation for relativistic electrons can be written as{\cite{andersson_damping_2001,stahl_synchrotron_2013,aleynikov_theory_2015}},
\begin{align}
\label{eq:runaway-fokker-planck}
  &\frac{\partial f}{\partial \hat{t}}+\frac{1}{p^{2}}\frac{\partial}{\partial p}\left[p^{2}\hat{E} f\right]+\frac{\partial}{\partial \xi}\left[\frac{1-\xi^{2}}{p}\hat{E}f\right]&\nonumber\\
  &-\frac{1}{p^2}\frac{\partial}{\partial p}\left[\left(1+p^{2}\right)f\right]-\frac{Z+1}{2}\frac{\sqrt{1+p^{2}}}{p^{3}}\frac{\partial}{\partial\xi}\left[(1-\xi^{2})\frac{\partial f}{\partial \xi}\right]&\nonumber\\
  &+\frac{1}{\hat{\tau}_r}\left\{-\frac{1}{p^{2}}\frac{\partial}{\partial p}\left[p^{3}\gamma (1-\xi^2) f\right]+\frac{\partial}{\partial \xi}\left[\frac{1}{\gamma}\xi(1-\xi^2) f\right]\right\}=0,&
\end{align}
where $p$ is the electron momentum (normalized to $m_{e}c$), $\xi$ is the cosine of the pitch angle, $Z$ is the ion effective charge, $\hat{E}=E/E_{c}$ where $E_{c}$ is the Connor-Hastie critical electric field $E_{c}=n_{e}e^{3}\ln\Lambda/\left(4\pi \epsilon_{0}^{2}m_{e}c^{2}\right)$ and $\ln\Lambda$ is the Coulomb logorithm, $\hat{t}=t/\tau$ where $\tau$ is the relativistic electron collision time $\tau=m_{e}c/\left(E_{c}e\right)$, $\hat{\tau}_{r}=\tau_{r}/\tau$ and $\tau_{r}$ is the timescale for the synchrotron radiation energy loss $\tau_{r}=6\pi \epsilon_{0} m_{e}^{3}c^{3}/\left(e^{4}B^{2}\right)$.

{In the adjoint method, we} define $P(p_{0}, \xi_{0})$ as the runaway probability function, which means the probability for an electron that is initially at $(p_{0},\xi_{0})$ to eventually run away. As shown in Appendix \ref{adjoint-equatin}, $P$ satisfies the homogeneous adjoint equation of Eq. (\ref{eq:runaway-fokker-planck}),
\begin{equation}
\label{eq:runaway-homo-adjoint}
  \mathcal{E}\left[P\right]+\mathcal{C}\left[P\right]+\mathcal{S}\left[P\right]+\mathcal{R}\left[P\right]=0,
\end{equation}
where
\begin{equation}
  \mathcal{E}\left[P\right]=\hat{E}\left[\xi\frac{\partial P}{\partial p}+\frac{1-\xi^{2}}{p}\frac{\partial P}{\partial \xi}\right],
\end{equation}
\begin{equation}
  \mathcal{C}\left[P\right]=-\frac{1+p^{2}}{p^{2}}\frac{\partial P}{\partial p},
\end{equation}
\begin{equation}
  \mathcal{S}\left[P\right]=\frac{Z+1}{2}\frac{\sqrt{1+p^{2}}}{p^{3}}\frac{\partial}{\partial\xi}\left[(1-\xi^{2})\frac{\partial P}{\partial \xi}\right]
\end{equation}
\begin{equation}
  \mathcal{R}\left[P\right]=\frac{1}{\hat{\tau}_r}\left[-\gamma p(1-\xi^2)\frac{\partial P}{\partial p}+\frac{1}{\gamma}\xi(1-\xi^2)\frac{\partial P}{\partial \xi}\right],
\end{equation}
{where the four terms represent, respectively, the parallel electric field force, the drag force in collision operator, the pitch angle scattering, and  the synchrotron radiation reaction force.}
The boundary conditions of $P$ are set as $P(p=p_{\mathrm{min}},\xi)=0,\quad P(p=p_{\mathrm{max}},\xi)=1$,
where $p_{\mathrm{min}}$ and $p_{\mathrm{max}}$ are two boundaries in momentum space that are located far from the transition region. (The solution is checked to be insensitive to the boundary locations)

\begin{figure}[ht]
  \includegraphics[width=0.5\textwidth,natwidth=61.74in,natheight=37.03in]{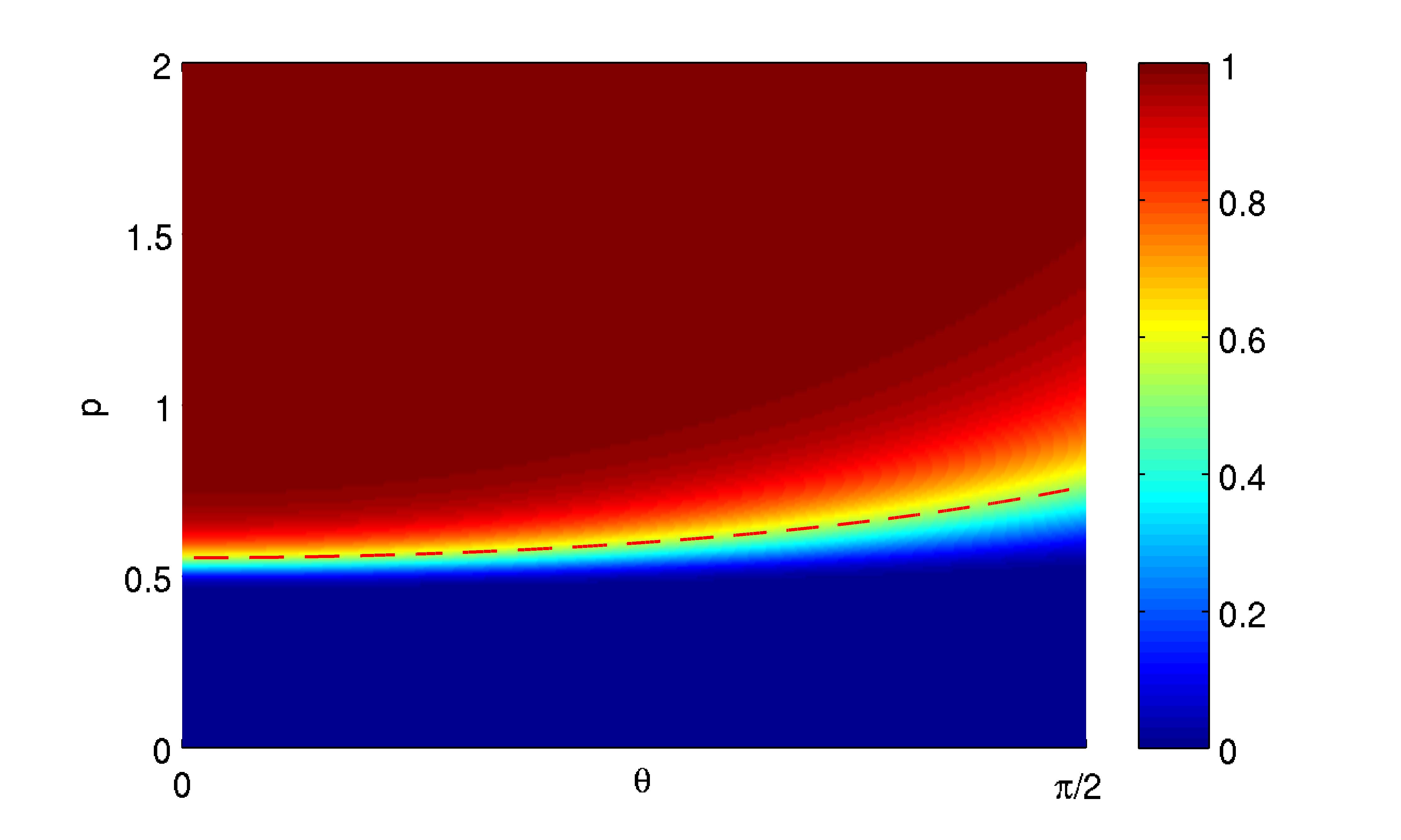}
  \caption{\label{runawayratio}The runaway probability for $E=6E_{c}$, $Z=1$ and $\hat{\tau}_{r}=100$. $\theta$ is the pitch angle. The red dashed line is the separatrix calculated using the test particle method in Ref. [\citenum{martin-solis_momentumspace_1998}].}
\end{figure}

We solve Eq. (\ref{eq:runaway-homo-adjoint}) numerically using the finite difference method, which is similar to the numerical method in Ref. \citenum{landreman_numerical_2014}. Figure \ref{runawayratio} shows $P$ for $E/E_{c}=6$, $Z=1$ and $\hat{\tau}_{r}=100$. The separatrix calculated using the test particle method in Ref. \citenum{martin-solis_momentumspace_1998} is also shown for reference. Note that the separatrix lies in the transition region of $P$ ($P$ between 0 and 1). However, we now have a smooth function that transtions from 0 to 1 rather than a step function. The width of the transition region depends on the amplitude of the pitch angle scattering term,  which increases with $Z$. This transition region is not captured in the test particle method.

\begin{figure}
  \includegraphics[width=0.4\textwidth]{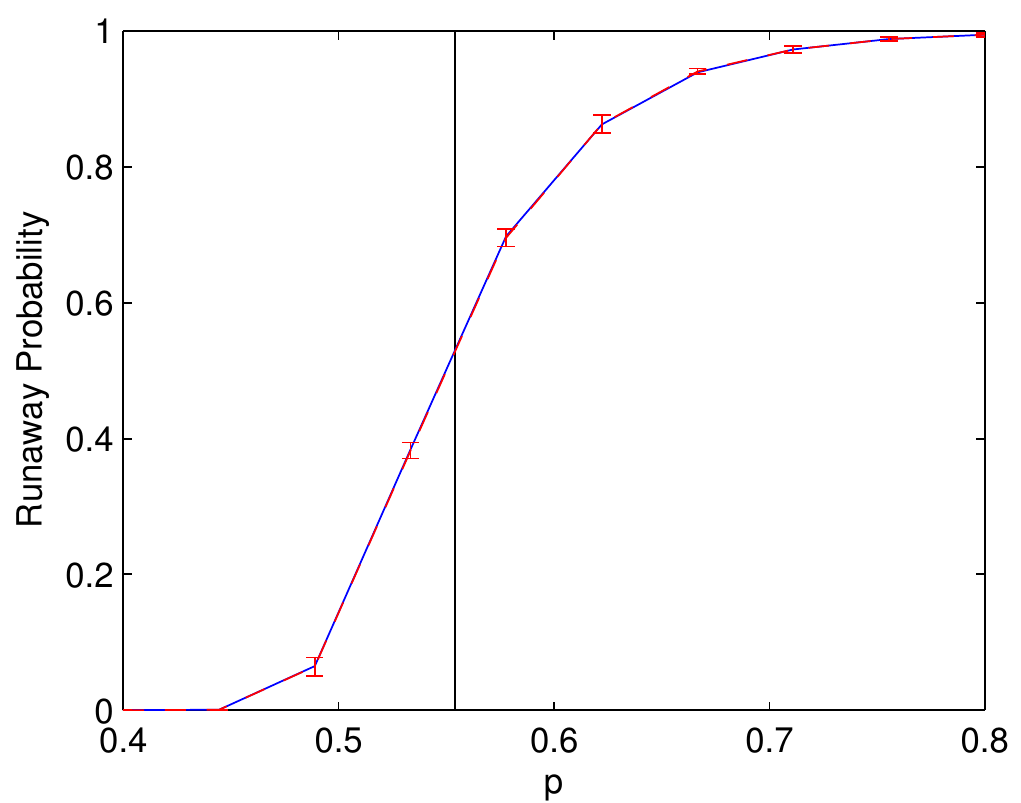}
  \caption{\label{monte-carlo}The runaway probability function $P$ for $\xi=1$ (blue line), compared with the Monte-Carlo simulation result (red dashed line) and the separatrix calculated by the test particle method (black vertical line). {The standard error of the Monte Carlo result is shown as error bars.}}
\end{figure}

Figure \ref{monte-carlo} shows the value of $P$ as a function of $p$ for $\xi=1$ in the transition region. The result is benchmarked with a Monte-Carlo simulation result, which is achieved by sampling a large number of electrons that start at one initial position and follow the equation of motion that corresponds to the FPE Eq. (\ref{eq:runaway-fokker-planck}). We then count the electrons that hit the low and high energy boundaries after a certain time. The two results are close. Note that unlike the Monte-Carlo method which can take significant computer time,  our method is fast and only requires solving the PDE once to obtain the probability function.\footnote{{Note that in the adjoint method we assume that physics quantities like $\hat{E}$ and $\hat{\tau}_{r}$ are time-independent. For a system with fast time-varying $E$ field or plasma density, the adjoint method is not applicable. However, in most cases including avalanche and plateau, the timescale for change in the electric field is much longer compared to the characteristic timescale for the movement of electrons in phase space, therefore the adjoint method is still applicable.}}

{$P$ has an increasingly sharp transition from near zero to near unity as $E$ increases to large values, which indicates that an electron with initial momentum above the separatrix} {is very likely to} {run away. However, as $E$ decreases, the value of $P$ above the separatrix} {reduces and eventually approaches} {zero.} {For $E$ sufficiently small} {$P$ becomes a flat function close to zero in most of momentum space, and only increases to 1 in a thin layer close to $p_{\mathrm{max}}$.\footnote{{This boundary layer effect is not physical but an artificial effect due to the choice of the high momentum boundary}} This is because the electric field is so weak that it is dominated by the combination of the collisional drag and the radiation force. Thus, all electrons, regardless of their initial energy, will slow down to the low energy regime in a finite time. This indicates the runaway population converts from growth {to decay.}

{In the decay phase the expected electron slowing-down time as a function of momentum, as opposed to the runaway probability, characterizes the decay of the runaway electron population.} {This can also be calculated using the adjoint method. Define $T(p,\xi)$ as the expected time for an electron initially at $(p,\xi)$ to reach the low energy boundary $p_{\mathrm{min}}$ or the high energy boundary $p_{\mathrm{max}}$. Note $1/T=1/T_{s}+1/T_{r}$, where $T_{s}$ is the expected slowing down time and $T_{r}$ is the expected time to run away. The ratio of the two terms is $(1-P)/P$. As shown in Appendix \ref{adjoint-equatin}, $T$ satisfies the nonhomogeneous adjoint FPE,}


\begin{equation}
\label{eq:nonhomo-runaway-adjoint}
  \mathcal{E}\left[T\right]+\mathcal{C}\left[T\right]+\mathcal{S}\left[T\right]+\mathcal{R}\left[T\right]=-1.
\end{equation}
The boundary conditions are $T(p=p_{\mathrm{min}},\xi)=0,\quad T(p=p_{\mathrm{max}}, \xi)=0$.
Note that for $E<E_{0}$, the runaway probability is close to zero almost everywhere, so $T_{r}\to\infty$ and $T\approx T_{s}$, except for the region near the high energy boundary.

Figure \ref{losttime} shows the calculated $T_{s}(p,\xi)$ for $\xi=1$ by solving Eq. (\ref{eq:nonhomo-runaway-adjoint}), for $\hat{E}=1.5$, $Z=1$. We see that $T_{s}$ is a monotonically increasing function of $p$. For small radiation force (large $\hat{\tau}_{r}$) and $E$ close to the critical field $E_{0}$, $T_{s}$ has a large jump between the low and high energy regimes. This is because in the intermediate energy regime all the forces reach a balance and the motion is dominated by the diffusion effect, therefore electrons take a very long time to cross this barrier region through random walk. For large radiation force and $E$ smaller than $E_{0}$ this jump is very small or non-existent because the radiation force is strong and always dominates the electric field force.

{We also calculate the effect of energy loss due to large angle collisions by including a Boltzmann collision operator using the {M\o ller} scattering cross section\cite{moller_zur_1932,boozer_theory_2015} in
 the adjoint equation. The result (dashed line in Figure \ref{losttime}) shows a significant decrease of slowing down time at the marginal case where $E$ close to $E_{0}$, while for $E$ much smaller than $E_{0}$ the result does not change much. This decrease occurs because in the marginal case, the diffusion barrier formed by the balance of forces is significant. The large angle collisions, however, can help electrons cross the barrier since they can cause a high energy electron to lose a large fraction of its energy and fall directly into the low energy regime. However, the jump in $T_{s}$ still exists.}
 
\begin{figure}
  \includegraphics[width=0.4\textwidth]{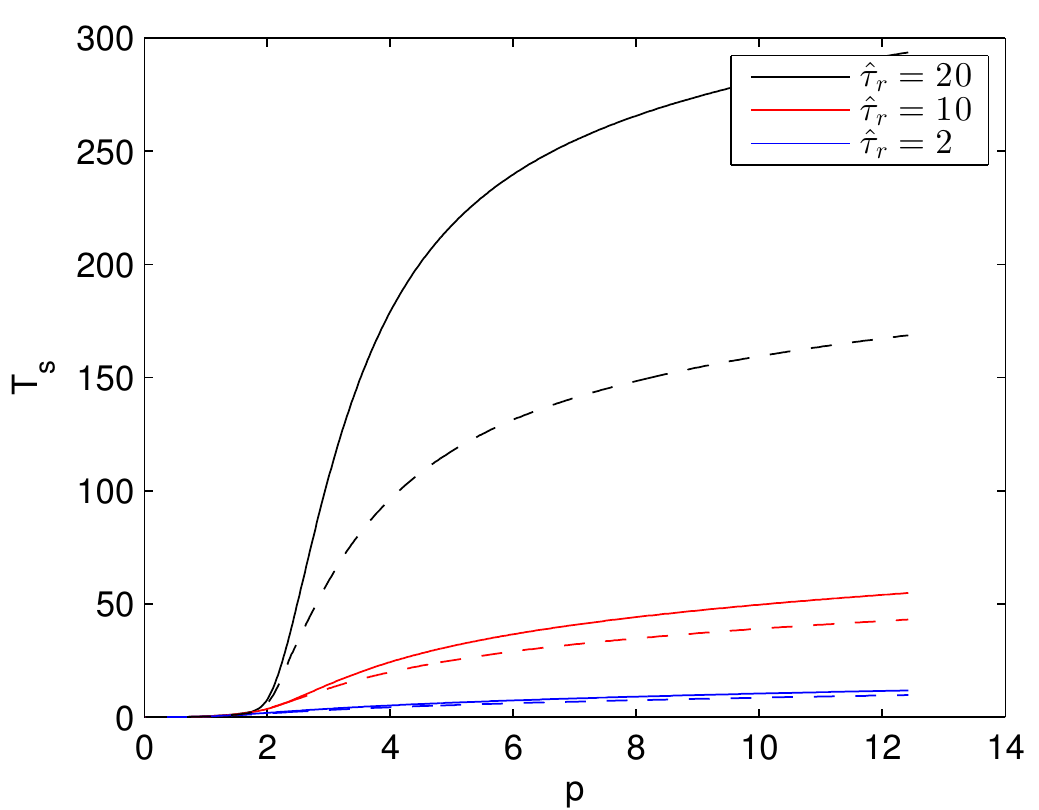}
  \caption{\label{losttime}The expected slowing-down time (normalized to $\tau$) as a function of $p$ at $\xi=1$ for $E/E_{c}=1.5$, $Z=1$ and 3 values of $\hat{\tau}_{r}$. Dashed line is the results including large angle collision effects with $\ln\Lambda=30$}
\end{figure}

The expected slowing-down time can be used to estimate the runaway electron beam decay time in experiments, and help explain the runaway electron population hysteresis and distribution. In both the QRE and current quench experiments, due to the decreasing magnitude of $E/E_{c}$, the runaway electron beam will have a transition from growth to decay. This means that at the beginning of the decay, there is already a population of high energy electrons formed by previous growth. The expected slowing-down time for these electrons determines the timescale for the runaway beam decay. In fact, if $E$ is very close to $E_{0}$, the expected slowing-down time for the high energy electrons can be very long, due to the jump in  $T_{s}$. This leads to a stagnation stage for the high energy electrons. The electric field required for a significant decay to happen is thus {far} lower than the critical electric field $E_{0}$, an effect first captured by this model. This can contribute to a hysteresis{\cite{aleynikov_theory_2015}} for the runaway electron population when the electric field is ramped up and down. {Another indication }{from the $T_{s}$ solution} {is that, due to very fast decay of the low energy electrons and extremely slow decay of the fast electrons,} {the electron population will} {tend to form a bump-on-tail distribution in the decay phase, which is different from the monotonic distribution in the growth {phase\cite{fulop_destabilization_2006}. Further investigations of the outcome of this distribution will be discussed in future work.}

{Returning to the} {runaway probability $P$}{, as $E$ is reduced $P$ will suddenly change} {from a smooth transition} {across the separatrix} {to a flat function} { near zero, indicating the cessation} {of the runaway} {generation} {process.} {This sudden change in the structure of $P$} {can be used to determine the critical electric field $E_{0}$ for runaway electron growth, which is above $E_{c}$ in the presence of the radiation force. One should bear in mind that this critical value is not an absolute threshold, given that $P$} {always smoothly transitions to 1 at $p_{\mathrm{max}}$, in a thin layer near $p_{\mathrm{max}}$ as $E$ approaches $E_{0}$.  We can, however, define a criterion based on the presence of a transition across the separatrix. }{For low $Z$ the} { transition} {is rather abrupt, while for high $Z$} {it} {is smoother.
Here we choose the} {a precise (but somewhat arbitrary) criterion that if $P$ is above 0.005 in the region above the separatrix, which means an electron there has a $0.5\%$ probability to run to the high energy boundary, then the runaway generation process is active.}
{We then} obtain $E_{0}$ from this {criterion}.

We have calculated $E_{0}$ for $1\le Z\le 10$  and $10\le \hat{\tau}_{r}\le 100${, as shown in Figure \ref{ERdifference}.} {The high energy boundary is chosen to be 30MeV.} A convenient function that fits the result is
\begin{equation}
  \frac{E_{0}}{E_{c}}=1+\alpha x^{\nu},
\end{equation}
\begin{equation}
  x=\frac{Z+1}{(\tau_{r}/\tau)^{3/4}},\quad \alpha=1.8587,\,\nu=0.6337.
\end{equation}
The error of the fitting function is less than 5\%.
\begin{figure}[ht]
  \includegraphics[width=0.4\textwidth]{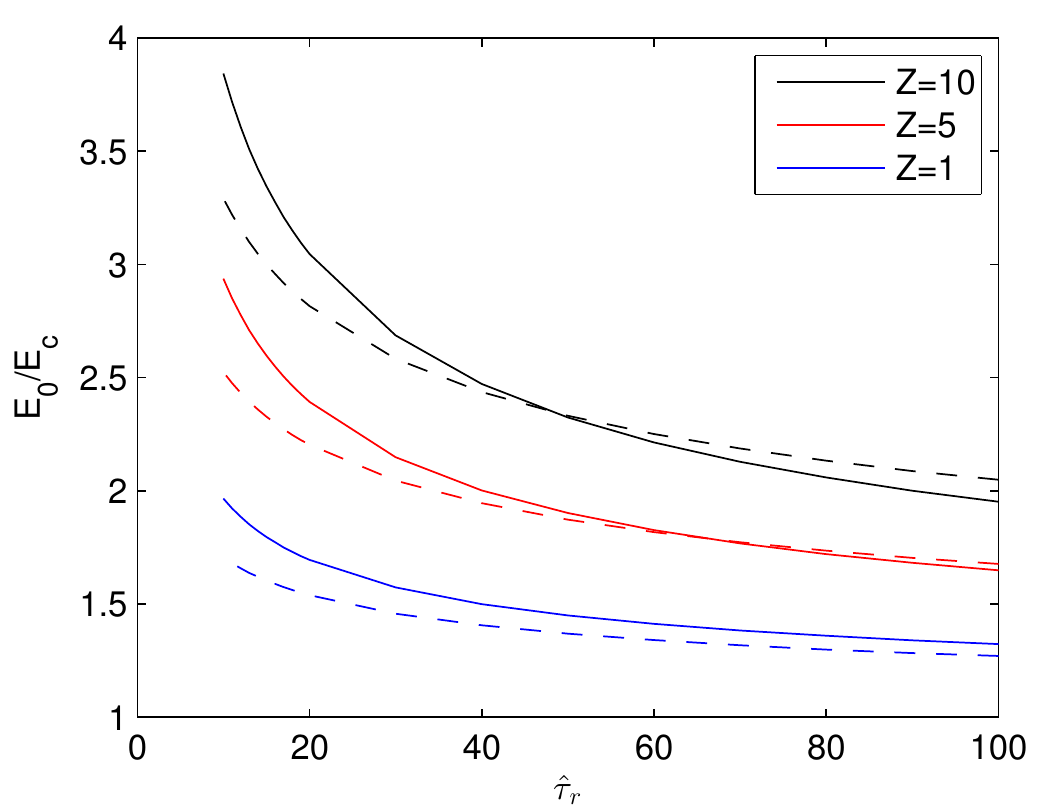}
\caption{\label{ERdifference}$E_{0}$ (solid line) plotted as a function of $\hat{\tau}_{r}$ for various $Z$. $E_{0}$ from Ref. \citenum{aleynikov_theory_2015} (dashed line) plotted for comparison.}
\end{figure}
We  {have also} benchmarked our result with $E_{0}$ in Ref. \citenum{aleynikov_theory_2015} in Figure \ref{ERdifference}.  The two results are close for small $Z$, while for large $Z$ our result is larger for small $\hat{\tau}_{r}$ but smaller for large $\hat{\tau}_{r}$.
{The discrepancy is mainly due to the difference in} {our definition of $E_{0}$} {and the uncertainty introduced by the smoother change of $P$ for high $Z$.}

Note that this critical electric field may be different from experimental observations for several reasons. If the electron temperature is very low or the pitch angle scattering is strong, the critical energy required for an electron to run away (the least momentum above the separatrix) is very high, which results in a growth rate too low to be observed. Additionally, the observed electric field corresponding to the turning point of the signal in the QRE experiments\cite{paz-soldan_growth_2014} can be higher than $E_{0}$, due to the energy dependence of the diagnostic and the redistribution of the runaway electron energy\cite{stahl_synchrotron_2013}.

{It is noteworthy that the the result of the adjoint method, especially the expected slowing-down time, depends highly on the energy diffusion mechanism in the kinetic model.} {In the results presented here,} {the model includes} {collisions and the} {synchrotron radiation reaction force. However,
the method can be easily extended to study the influence of other physics on runaway electron dynamics, including  Bremsstrahlung radiation\cite{fernandez-gomez_determination_2007} and magnetic fluctuations\cite{martin-solis_effect_1999}, by adding the corresponding operators into} {the} {adjoint FPE. }In addition, the adjoint {FPE} can be used to study any dynamical system that has a separatrix or a singular point, e.g. particle behavior close to the magnetic separatrix and the X-point. 
Future applications of this method to other areas are promising.

We thank O. Embr\'{e}us, I. Fern\'{a}ndez-G\'{o}mez, N. Fisch, T. F\"{u}l\"{o}p, P. Helander, E. Hirvijoki, J. Krommes, G. Papp and A. Stahl for useful discussions. The numerical calculations are conducted on the PPPL Beowulf cluster.
This work is supported by the U.S. Department of Energy under Contract No. De-FG02-03ER54696.

\appendix
\label{adjoint-equatin}
\emph{Appendix: Adjoint Fokker-Planck equation.} --- Here we introduce the adjoint FPE method{, begining with} the homogeneous adjoint FPE. Consider a test particle in a stationary stochastic system. Denote the particle's coordinate as $x$ with two boundaries $x_{\mathrm{min}}$ and $x_{\mathrm{max}}$.
The equation of motion of the test particle in the stochastic system can be described as $\delta x=x(t+\delta t)-x(t)=v(x)\delta t+\xi(x)$, where $\xi(x)$ is a random variable that satisfies $\langle\xi(x)\xi(x)\rangle=\sqrt{2 D(x) \delta t}$.
The distribution function $f(x,t)$ thus satisfies the FPE
\begin{equation}
\label{eq:fokker-planck}
  \frac{\partial f}{\partial t}=-\frac{\partial}{\partial x}\left[v(x) f \right]+\frac{\partial^{2}}{\partial x^{2}}\left[D(x) f \right].
\end{equation}

Define $P(x_{0})$ as the probability of a test particle with initial coordinate $x=x_{0}$ to first pass the boundary $x_{\mathrm{max}}$ rather than $x_{\mathrm{min}}$. Note that because the system is stationary, $P$ is time-independent. $P(x_{0})$ can be expressed using the random walk probability density in terms of the particle's next-step coordinate,
\begin{align}
\label{eq:next-step}
  P(x_0)&=\int P(x_0+\delta x) G(x_{0}, \delta x)  d\delta x\nonumber\\
  &=\int \left[P(x_0)+\frac{d P(x_0)}{dx}\delta x+\frac{1}{2}\frac{d P(x_0)}{dx^{2}}\delta x^{2}\right]\nonumber\\
  &\qquad \quad G(x_{0},\delta x)d\delta x,
\end{align}
where $G(x_{0}, \delta x)$ is the probability density that the coordinate can have a change of $\delta x$ in $\delta t$ if $x=x_{0}$, and we expand in powers of $\delta x$ in anticipation of taking the limit $\delta x\to 0$. $G(x_{0},\delta x)$ satisfies the following properties
\begin{equation}
  \int G(x_{0}, \delta x) d\delta x=1,\quad\int G(x_{0},\delta x)\delta x d\delta x=v(x_{0}) \delta t,
\end{equation}
\begin{equation}
  \int G(x_{0},\delta x)\delta x^{2} \delta x=v(x_{0})^{2}\delta t^{2}+2D(x_{0})\delta t.
\end{equation}
Using these equations, Eq. (\ref{eq:next-step}), and taking the limit $\delta t\to 0$, we obtain the differential equation for $P(x)$,
\begin{equation}
\label{eq:adjoint-fokker-planck}
  v(x)\frac{d P(x)}{dx}+D(x)\frac{d^{2}P(x)}{dx^{2}}=0,
\end{equation}
which is the adjoint equation of Eq. (\ref{eq:fokker-planck}).
According to the definition, the boundary conditions obeyed by $P(x)$ are $P(x=x_{\mathrm{min}})=0,\qquad P(x=x_{\mathrm{max}})=1$. The probability for the particle to first pass the boundary at $x_{\mathrm{min}}$ can be obtained simply from the relation $Q(x_{0})=1-P(x_{0})$. Note that $Q$ also satisfies Eq. (\ref{eq:adjoint-fokker-planck}).

We next discuss the nonhomogeneous adjoint Fokker-Planck equation. Define $T(x_{0})$ as the expected time for a test particle that starts at $x=x_{0}$ to first pass either boundary, $x_{\mathrm{min}}$ or $x_{\mathrm{max}}$. Similar to $P$, $T$ can also be calculated through the random walk integral,
\begin{align}
\label{eq:next-step2}
  T(x_0)&=\int T(x_0+\delta x) G(x_{0}, \delta x) d\delta x+\delta t\nonumber\\
  &=\int \left[T(x_0)+\frac{d T(x_0)}{dx}\delta x+\frac{1}{2}\frac{d T(x_0)}{dx^{2}}\delta x^{2}\right]\nonumber\\
  &\qquad\quad G(x_{0},\delta x) d\delta x+\delta t.
\end{align}
Taking the limit $\delta t\to 0$, $T(x)$ is found to satisfy the differential equation
\begin{equation}
\label{eq:adjoint-fokker-planck2}
  v(x)\frac{d T(x)}{dx}+D(x)\frac{d^{2}T(x)}{dx^{2}}=-1,
\end{equation}
which is analogous to Eq. (\ref{eq:adjoint-fokker-planck}) except the equation is now nonhomogeneous. The boundary conditions for $T$ are simply $T(x=x_{\mathrm{min}})=0,\qquad T(x=x_{\mathrm{max}})=0$.

Let us assume a particle source at $x=x_{0}$. The rate for particles to pass one of the boundaries can be expressed as $r=1/T$. Note that $r=r_{1}+r_{2}$, where $r_{1}$ and $r_{2}$ are the rate to pass the boundary at $x_{\mathrm{min}}$ and $x_{\mathrm{max}}$. Both $r_{1}$ and $r_{2}$ can then be calculated according to the first passage probability, $r_{1}/r_{2}=Q/P$.

{It is important to point out that, though the derivation shown here is based simply on finding the adjoint FPE, the adjoint method can also be applied to more general kinetic PDEs. For example, to treat large angle collision effects one needs to use the Boltzmann collision operator, in which case the kinetic equation is an integro-differential equation rather than a differential one. However, one can still find the adjoint equation through integration by parts, or from the first line in Eq. (\ref{eq:next-step}) and Eq. (\ref{eq:next-step2}) without performing Taylor expansion.}


\bibliography{Adjoint}
 
\end{document}